\documentclass[journal]{IEEEtran}
\ifCLASSINFOpdf

\else

\fi

\usepackage{graphics}

\usepackage{amssymb}
\usepackage{amsmath}
\usepackage{amsthm}

\usepackage{pdflscape}
\usepackage{rotating}
\usepackage{gensymb}
\usepackage{amssymb}
\usepackage{pifont}
\newcommand{\cmark}{\ding{51}}%
\newcommand{\xmark}{\ding{55}}%

\DeclareGraphicsExtensions{.png}

\usepackage{color}

\newtheorem*{mydef}{Definition}

\begin{document}
%
\title{moGrams: A Network-based Methodology for Visualizing the ​​Set of Non-dominated Solutions ​in Multiobjective Optimization}

\author{Krzysztof~Trawi\'nski,
	Manuel~Chica,
	David P. Pancho,
	Sergio~Damas, 
        and~Oscar~Cord\'on
\thanks{Krzysztof~Trawi\'nski, Manuel~Chica, David~P.~Pancho, Sergio~Damas, and Oscar Cord\'on are with European Centre for Soft Computing, 
Edificio Cient\'{\i}fico-Tecnol\'ogico, planta 3, C. Gonzalo Guti\'errez Quir\'os s/n,
33600 - Mieres (Asturias), SPAIN
e-mail: \{krzysztof.trawinski, manuel.chica, david.perez, sergio.damas, oscar.cordon\}@softcomputing.es.}
\thanks{Oscar Cord\'on is also with the Dept. of Computer Science and Artificial Intelligence (DECSAI) and
the Research Center on Information and Communication Technologies (CITIC-UGR),
University of Granada, 18071 Granada, SPAIN
e-mail: ocordon@decsai.ugr.es}}

\markboth{}%
{Shell \MakeLowercase{\textit{et al.}}: Bare Demo of IEEEtran.cls for Journals}
%



\maketitle

\begin{abstract}
An appropriate visualization of multiobjective non-dominated solutions is a valuable asset for decision making. Although there are methods for visualizing the solutions in the design space, they do not provide any information about their relationship. In this work, we propose a novel methodology that allows the visualization of the non-dominated solutions in the design space and their relationships by means of a network. The nodes represent the solutions in the objective space, while the edges show the relationships between the solutions in the design space. Our proposal (called \emph{moGrams}) thus provides a joint visualization of both objective and design spaces. It aims at helping the decision maker to get more understanding of the problem so that (s)he can choose the more appropriate final solution. \emph{moGrams} can be applied to any multicriteria problem in which the solutions are related by a similarity metric. Besides, the decision maker interaction is facilitated by modifying the network based on the current preferences to obtain a clearer view. An exhaustive experimental study is performed using three multiobjective problems in order to show both the usefulness and versatility of \emph{moGrams}. The results exhibit interesting characteristics of our methodology for visualizing and analyzing solutions of multiobjective problems.
\end{abstract}


%
\IEEEpeerreviewmaketitle


\section{Introduction}
\label{sec:intro}

Most of the real-world problems have a multicriteria nature (i.e. they include several conflicting criteria)~\cite{Chankong83}. Multiobjective (MO) optimization methods are designed for solving and produce a set of non-dominated solutions~\cite{Coello07,Deb01} when tackling these problems. Decision makers (DMs) aim at selecting the best possible solution from the set of returned alternatives. To do so, the DM has to compare the alternative solutions and that is a difficult and time-consuming task. 
In evolutionary multiobjective optimization (EMO)~\cite{Coello07,Deb01} some authors define a global framework considering multicriteria decision making (MCDM) as a conjunction of three components: search, preference trade-offs, and visualization~\cite{Bonissone08}.

A helpful visualization process in EMO should provide a deeper understanding of the problem and new useful information regarding the alternatives. The DM should receive as much relevant information as possible provided by powerful methods and techniques (e.g. networks and other visualization tools). A good visualization enables her/him to obtain more insights of the problem and the different solutions to identify differences and similarities before coming to the final decision~\cite{Miettinen14}. In particular, the flexibility (i.e., the ease to change one solution by another in the decision space) is an important property to respond to frequent environmental changes in many managerial and operation research problems in which the information is uncertain~\cite{Chica16OMEGA}. Additional information about the flexibility of the non-dominated solutions will be really worthy for the DM.

There is an increasing number of studies demonstrating that visualization combined with optimization can promote design innovations and provide DMs with an improved understanding of the problem~\cite{Fleming05,Stump03}. The visualization of solutions in MO optimization is an active research area~\cite{Walker13,Tusar15} and most of the works are devoted to visualization of the Pareto front in the objective space (see for example~\cite{Lotov04,Kollat07,Kurasova13}). Recent studies focus on how to represent non-dominated solutions for many-objective problems (i.e. those having more than three objectives)~\cite{Walker13}.
However, just a few proposals deal with the visualization of the MO solution in the design space (e.g. self organization maps~\cite{Obayashi05}, heatmaps~\cite{Pryke07}, cloud visualization~\cite{Eddy02} or hyper-space diagonal counting~\cite{Agrawal04}). 
Up to our knowledge, there has not been any previous proposal providing the DM with insights about the relationships between the solutions in the decision space with the aim of helping the decisor to understand the problem and assisting to choose the final solution.


In this paper, we aim at proposing a novel methodology, called \emph{moGrams}, for visualizing and analyzing MO solutions in order to facilitate the MCDM process. Our novel framework provides a clear insight of the design space by showing the relations among the obtained MO solutions in a network representation. Moreover, \emph{moGrams} provide objective space information of the solutions in a joint visualization of both design and objective spaces.

A \emph{moGram} is a weighted network where its nodes represent non-dominated solutions and its edges represent design space relations between the solutions. A similarity metric is used to generate the weighted edges of the network. This metric is defined by the DM and is specific to the MO problem. This is thus a generic methodology that applies to any multicriteria problem where its solutions can be related by a similarity metric. Also, \emph{moGrams} deal with the scalability problem as multiple solutions with many objectives are shown in a salient way. In particular, every node (a non-dominated solution) is splitted into a number of sectors that corresponds to the number of objectives of the MO problem. 
 
Our network-based proposal also enhances the similarity analysis of the solutions in the decision space. The analysis facilitates the evaluation of the flexibility of every solution. For instance, the DM would prefer a very connected node within the network. This is because that solution (represented by the node) could be easily transformed into an adjacent solution (high similarity in the design space). 

\emph{moGrams} use social network analysis (SNA) techniques to improve the visual information from the network~\cite{Scott2000,wasserman1994social}.
In particular, the Pathfinder algorithm~\cite{Dearholt90,Sch89} reduces the complexity of the network by only representing the most salient connections, making its analysis easier. 
In order to clearly visualize the network, a force-directed layout is used, namely the Kamada-Kawai algorithm~\cite{MoyaAnegon07}. \emph{moGrams} is also an interactive visualization framework where the DM has the possibility to obtain a clearer view of the network according to her/his preferences. 

In order to show the benefits and versatility of \emph{moGrams} we perform a complete experimentation on three MO problems. These three MO application cases come from different research fields and we choose them because of their different features (e.g. their decision variables are very diverse, from combinatorial optimization to real parameters). The first MO application case is the time and space assembly line balancing problem (TSALBP)~\cite{Chica10Ins,Chica11CAIE} which deals with the joint optimization of the number of stations and their area when configuring an industrial line for assembling products. The next one considers the overproduce-and-choose strategy (OCS)~\cite{Partridge96} for classifier ensembles (CEs)~\cite{Kun04}and it aims at obtaining CEs with a low number of base classifiers, keeping a good accuracy~\cite{Trawinski11,Trawinski2013}. The last application case focuses on the inductive query by example (IQBE)~\cite{Cordon2006}. IQBE is a paradigm that allows a system to automatically derive queries for a specific information retrieval system, defined as a MO problem with a conflicting precision-recall trade-off.

The rest of the paper is structured as follows. Section~\ref{sec:background} reviews the state of the art in MO visualization and a background of SNA. Section~\ref{sec:mograms} provides a description of our novel visualization proposal. In order to show usefulness of \emph{moGrams}, Sections~\ref{sec:appcase1},~\ref{sec:appcase2} and~\ref{sec:appcase3} present three application cases including TSALBP, CEs and IQBE, respectively. Finally, Section~\ref{conclusions} outlines concluding remarks and future research works.



\section{Background}
\label{sec:background}

\subsection{Visualization of Multicriteria Solutions}
\label{sec:soa}

There are multiple proposals for visualizing the MO and many-objective Pareto front solutions in the literature~\cite{Walker13}. Probably, the most common method for visualizing MO solutions is a scatterplot~\cite{Chambers82}. One of the most popular choices for many-objective solutions is the use of parallel coordinate plots \cite{inselberg09}.
Recent reviews~\cite{Walker13,Tusar15} show that most of the works in the literature are devoted to the visualization of MO Pareto front solutions in the objective space. 

The visualization of multicriteria solutions in the design space did not receive much attention indeed. 
%
We present a summary of the reviewed proposals in Table~\ref{table:soa}. First and second columns include the reference and visualization method used, respectively. The third column indicates the additional algorithms used to either obtain a clearer visualization (e.g. hierarchical clustering, spectral seriation) or provide additional numerical information (e.g. ANOVA, data mining classification). Only some proposals considered a joint visualization for objective and design spaces (fourth column). None of these works were focused on the similarities between the solutions in the design problem space (penultimate column) and none of them used a network as a tool for visualization of the MO solutions (last column).

\begin{table*}[htb!] 
\caption{Publications considering the design space visualization of MO solutions.} \label{table:soa}
\centering
\large
\scalebox{0.6}{
\begin{tabular}{|cccccc|}
 \hline
  \textbf{Reference(s)} & \textbf{Visualization method} & \textbf{Supplementary algorithms} & \textbf{Joint} & \textbf{Relations between} & \textbf{Network-based} \\
 & & & \textbf{visualization} & \textbf{solutions in decision space} & \textbf{visualization} \\
 \hline 
  Eddy and Lewis~\cite{Eddy02} & Cloud visualization & - & \xmark & \xmark & \xmark  \\
  Winer and Bloebaum~\cite{Winer02} & Graph morphing & - & \cmark & \xmark & \xmark  \\
  Agrawal et al.~\cite{Agrawal04,Agrawal06} & HSDC & - & \cmark & \xmark & \xmark  \\
  Obayashi et al.~\cite{Obayashi05,Jeong05} & SOM & ANOVA & \cmark & \xmark & \xmark  \\
  Pryke et al.~\cite{Pryke07} & Heatmap & Hierarchical clustering & \cmark & \xmark & \xmark  \\
  Jeong et al.~\cite{Obayashi05,Jeong07} & Synchronous visualization & - & \xmark & \xmark & \xmark  \\
  Kollat and Reed~\cite{Kollat07} & Spatial coordinates, size, shape, color, etc. & - & \xmark & \xmark & \xmark  \\
  Blasco et al.~\cite{Blasco08} & Level diagrams & Data mining classification & \xmark & \xmark & \xmark  \\ 
  Walker et al.~\cite{Walker13} & Heatmap &  Spectral seriation & \cmark & \xmark & \xmark  \\
  Kubota et al.~\cite{Kubota14} & Scatterplot and parallel coordinate plot & - & \xmark & \xmark & \xmark  \\
 \hline
\end{tabular}
}
\end{table*}

Self-organizing maps (SOMs) attracted attention as a novel means for visualization of both objective and design space. 
SOM projects multidimensional data on a 2-D map without any information loss.
In~\cite{Obayashi05,Jeong05}, SOM was applied to find the trade-offs between objective spaces, relationships between objective spaces and design variables. 
%
%
In both contributions, analysis of variance, namely ANOVA, was used to show the effect of each design variables on objective functions in a quantitative way.

In a different manner, Pryke et al. introduced in~\cite{Pryke07} a heatmap for presenting both the objective and parameter space in the same view. A hierarchical clustering was used to alleviate the transparency issue. Walker et al.~\cite{Walker13} also used heatmaps for MO visualization and improved their readability by using spectral seriation which rearranges the objective or parameter space views on the heatmap. 

Agrawal et al.~\cite{Agrawal04,Agrawal06} presented an intuitive visualization methodology for multidimensional MO optimization problems by using the hyper-space diagonal counting which enables a lossless mapping of dimensions. The proposed method dealed with more than three objectives and also with design space visualization.


There are some proposals providing a visualization of both objective and decision spaces in separate windows such as cloud visualization~\cite{Eddy02} and synchronous visualization~\cite{Jeong07}. The visualization frameworks, VIDEO~\cite{Kollat07} and EVOLVE~\cite{Kubota14}, followed the same idea. They used standard techniques such as spatial coordinate axes, color, Kriging mapping in VIDEO; and scatterplot with Parallel Coordinate Plot in EVOLVE.



Besides, we can highlight other interesting proposals where authors presented each objective and design variable in separate windows (level diagrams)~\cite{Blasco08} or showed several three dimensional visualizations with fixed values of some design variables (graph morphing)~\cite{Winer02}.





\subsection{Social network analysis}
\label{sec:sna}

The use of SNA techniques has proved its capability to achieve high quality, schematic visualizations of the resulting network-based representations in various fields: psychology (to represent the cognitive structure of a subject~\cite{Dearholt90,Sch89}), system behavior (for designing and analyzing fuzzy systems~\cite{PanchoIEEETFS2013}), scientometrics (for the analysis of large scientific domains~\cite{Vargas2007visualizing}). Among others, SNA techniques are especially useful to solve two tasks:

\subsubsection{Network scaling}
Networks are usually dense and scaling is necessary to obtain structures revealing the underlying organization, maintaining all the nodes but only the most important relations. Three predominant SNA alternatives to accomplish this task are presented in the literature~\cite{chen2003visualizing}: The first method discards edges with weights below a given threshold~\cite{zizi1994accessing}. This approach, although easy to implement, does not consider intrinsic structure of the underlying network. Thus, the transformed network may not show the nature of the original one. 
The second method extracts a minimum spanning tree of the network~\cite{noel2002visualization}. This guarantees a fixed number of edges (a number of nodes minus one), although 
it may not show the underlying information. The last method provides constraints on paths and removes the edges not satisfying them. The Pathfinder algorithm~\cite{Dearholt90,Sch89} is one of the most popular, known for its mathematical properties associated to preservation of the triangular inequality. Some of these properties are the conservation of the edges, the capability of modeling symmetrical and also asymmetrical relationships, maintenance of sub-networks, as well as the representation of the most salient relationships from the data.

\subsubsection{Network drawing}
There are different SNA methods for automatic visualization of networks. Force-based or force-directed algorithms are the most widely used algorithms for drawing networks in the area of information science~\cite{battista1999graph,kobourov2005force}. Their purpose is to locate the nodes of a network in a two or three dimensional space so that all the edges are approximately of equal length and there are as few crossing edges as possible, trying to obtain the most aesthetically pleasing view. Kamada-Kawai~\cite{KamadaKawai1989} and Fruchterman-Reingold~\cite{fruchterman1991graph} are their most representative methods of the network drawing family.



\section{Our visualization proposal}
\label{sec:mograms}

In this section, we present our visualization framework. We first define and describe the generation of \emph{moGrams} in Sections~\ref{sec:mogramsDef} and~\ref{sec:mogramsGen}, respectively. Then, we show DM implications in Section~\ref{sec:dmimplic}.

\subsection{moGrams definition}
\label{sec:mogramsDef}

A \emph{moGram} is a network, that jointly presents the solutions of a MO problem in the objective space and their relationship in the design space. In particular, every node of the network represents a non-dominated solution in the objective space and its edges represent design space relations between the solutions. There are as many nodes as non-dominated solutions in the objective space. The formal definition of the \emph{moGram} is as follows:

\begin{mydef}
 moGram \emph{is defined as a tuple} $G=(N,A)$ \emph{where} $N$ \emph{is the set of nodes representing the $n$ non-dominated solutions and} $A$ \emph{is the adjacency matrix.} $A(i,j)=1$ \emph{means there is an edge between nodes $i$ and $j$ and therefore, there is a significant relationship in the design space between the solutions represented by those nodes, according to the similarity metric.}
\end{mydef}


\subsection{moGrams generation}
\label{sec:mogramsGen}



\begin{figure*}[htb!]
\begin{center}
\includegraphics[width=13cm]{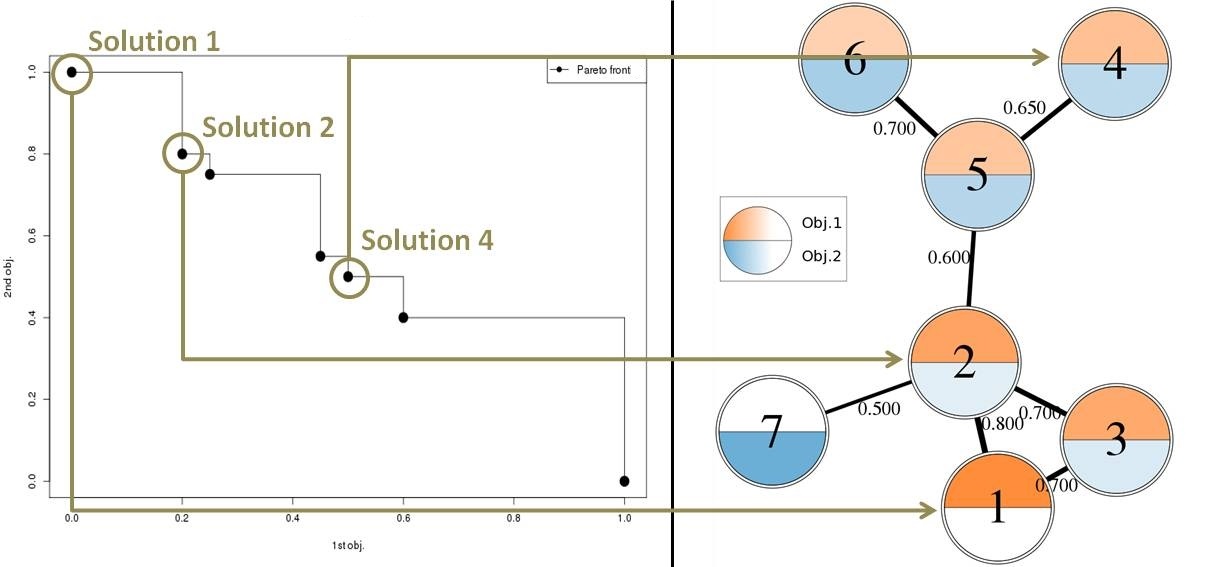}
\caption{A Pareto front approximation (on the left) and the corresponding \emph{moGram} (on the right) for a toy example.}
\label{fig:toyEx}
\end{center}
\end{figure*}

To facilitate comprehension of the \emph{moGrams} generation, we use an illustrative example presented in Fig.~\ref{fig:toyEx}. It shows the scatterplot visualization of the Pareto front approximation on the left-hand side and generated \emph{moGram} on the right-hand side. The illustrative example is a minimization problem consisting of 7 non-dominated solutions with two objectives and a similarity metric for the design space. 

The generation of a \emph{moGram} is divided into the following four phases:
\begin{enumerate}

 \item \emph{Design space visualization}. We generate a complete network with weighted edges based on $S$ (see Equation~\ref{eq:matrix}), a $n\times n$ symmetric matrix which contains the similarity values for all the non-dominated solutions. To generate $S$ we use the similarity metric $Sim(a, b) \in [0,1]$ which is a function returning a value for the closeness between two solutions $a$ and $b$ in the design space. This similarity metric is defined by the DM and is specific for the MO problem. We present a numerical example in Equation~\ref{eq:matToyEx}, which is a similarity matrix of the illustrative example (the most important edges are presented in bold font).
 

\begin{equation}
\label{eq:matrix}
S = 
\left(
\begin{array}{cccc}
1 & s_{12} & \dots & s_{1n} \\
s_{21} & 1 & \dots & s_{2n} \\
\vdots & \vdots & \ddots & \vdots \\
s_{n1} & s_{n2} & \dots & 1 \\
\end{array}
\right)
\end{equation}

\begin{equation}
\label{eq:matToyEx}
S_{toy} = 
\left(
\begin{array}{ccccccc}
1 & \textbf{0.8} & \textbf{0.7} & 0.1 & 0.1 & 0.1 & 0.1 \\
\textbf{0.8} & 1 & \textbf{0.7} & 0.1 & \textbf{0.6} & 0.1 & \textbf{0.5} \\
\textbf{0.7} & \textbf{0.7} & 1 & 0.1 & 0.1 & 0.1 & 0.1 \\
0.1 & 0.1 & 0.1 & 1 & \textbf{0.65} & 0.1 & 0.1 \\
0.1 & \textbf{0.6} & 0.1 & \textbf{0.65} & 1 & \textbf{0.7} & 0.1 \\
0.1 & 0.1 & 0.1 & 0.1 & \textbf{0.7} & 1 & 0.1 \\
0.1 & \textbf{0.5} & 0.1 & 0.1 & 0.1 & 0.1 & 1 \\
\end{array}
\right)
\end{equation}

The edge weight is proportional to the similarity of the two solutions and a label indicating its value is drawn in the network. It is also illustrated by the thickness of the edge to highlight it in the visual representation. For instance, in the example from Fig.~\ref{fig:toyEx}, it can be noticed that the edge between node 1 and 2 is thicker than the edge between node 2 and 7.

 \item \emph{Objective space visualization}. \emph{moGrams} provide information regarding the objective space of the solutions obtained by using different colors and opacity. 
The number of nodes of the \emph{moGram} corresponds to the number of non-dominated solutions. Every node is divided into $n_{obj}$ sectors of the same size, where $n_{obj}$ is the number of the objectives. One color is assigned to each objective whose transparency is proportional to the objective value of the given solution. Thus, \emph{moGrams} allow the DM to get a detailed insight of the solution not only in the design space but also in the objective space. 
 In the illustrative example (Fig.~\ref{fig:toyEx}), each node is divided in half (180\degree slices). We assign the orange color to the first objective (the upper half of the node), while the blue color represents the second objective (the lower half of the node). For example, solution 1 has a very strong orange color in the upper half and is white (the blue color is fully transparent) in the lower half. Thus, we can deduce that this node has a low value of the objective 1 and a high value of the objective 2, laying in the left upper extreme of the Pareto front approximation (as seen in the visualization of the Pareto front approximation). 

 \item \emph{Network scaling}. The resulting complete network is usually unreadable due to the large number of edges in most of the cases. Thus, the Pathfinder scaling method~\cite{Dearholt90,Sch89} is applied to $S$ in order to reduce the network and maintain only the most significant edges. 
By doing so, Pathfinder also generates the adjacency matrix $A$ of the network. As already said, the Pathfinder algorithm, has two important characteristics: 1) it retains the most important edges and 2) it does not produce the isolated nodes. Notice that, this phase deals with the scalability of the problem, as Pathfinder can deal with large networks and strongly reduces the number of connections between the nodes. Its results are demonstrated in Fig.~\ref{fig:toyEx}, where Pathfinder reduces the number of edges from 21 (the complete network) to only 7. 

 \item \emph{Network drawing}. Force-based algorithms are suitable to graphically visualize networks, as already mentioned in Section~\ref{sec:sna}. We use the Kamada-Kawai~\footnote{We use the Graphviz implementation of this algorithm. http://www.graphviz.org \cite{Graphviz1999}} algorithm, since it has been shown to work effectively when combined with Pathfinder for other problems such as system behavior~\cite{PanchoIEEETFS2013} and scientometrics~\cite{MoyaAnegon07}. Again, taking a look at the illustrative example, we can observe a salient and clear view of the solutions.

\end{enumerate}

\subsection{DM implications}
\label{sec:dmimplic}
%

\emph{moGrams} allow a DM a certain interaction with the network generated in order to get more insights of the given problem and the different solutions to be compared. For example, DM can adjust the similarities between the nodes. When their values belong to a certain narrow range, all edges in a \emph{moGram} will have almost the same thickness. By scaling the range of similarities, the modified~\emph{moGram} will provide a clearer view of the relationship between the nodes. Moreover, only the nodes with specific characteristics can be interesting for the DM. Thus, our framework allows to remove some nodes from the visualization. To accomplish that, the \emph{moGram} generation has to be relaunched excluding the marked nodes. The DM can also remove the information regarding the objective space, if the decisor is only interested in having a view into the design space. In this case, all nodes will have one uniform color only.



\section{Application case 1: assembly line balancing}
\label{sec:appcase1}

\subsection{Problem description}

An assembly line consists of a set of $m$ workstations and $n$ different tasks, all of them requiring an operation time for their execution. These tasks divide the manufacturing of a production item. One usual and difficult problem is to determine how these $n$ tasks can be assigned to $m$ stations fulfilling certain restrictions (assembly line balancing (ALB)~\cite{Battaia13}). The goal of the TSALBP, a family of ALB problems, is to optimally assign tasks to stations with respect to some objectives (cycle time of the line or the number of stations and/or their area) in such a way that all the precedence, time, and/or spatial constraints are satisfied~\cite{Chica10Ins}. This assignment is called assembly line configuration and it is a solution for the problem.

One of the TSALBP variants is the multiobjective TSALBP-m/A which tries to jointly minimize the number of stations $m$ of the assembly line and their line area ($A$) for a given product cycle time. This variant is a complex and realistic multicriteria problem in the automotive industry which favored the application of MO methods to solve it such as multiobjective ant colony optimization~\cite{Chica10Ins}, EMO~\cite{Chica11CAIE}, and memetic algorithms~\cite{Chica12EngAppAI}. 

All the latter methods are able to return a set of non-dominated solutions for a known demand of homogeneous products. However, as pointed out in~\cite{Chica13IJPE, Chica16OMEGA}, flexibility is an important asset to manufacturing firms to respond to changes in the environment and this flexibility also applies to the automotive industry and ALB. Providing DMs with additional information about how flexible a non-dominated solution is, would be valuable when making her/his decision. This flexibility can be seen as the number of changes to perform when moving from one solution to another in the decision space (measured by a similarity index between both solutions) also taking into account how  objective values change.


Therefore, we can apply \emph{moGrams} to the non-dominated solutions of the TSALBP in order to provide visual information about the difficulty involved in replacing one solution by another. Flexible solutions will facilitate that transition. Two TSALBP solutions (assembly line configurations) $\psi_1$ and $\psi_2$ are characterized by the assignment of $n$ tasks to $m_1$ and $m_2$ workstations, respectively, and their station workloads are $(S_1^{\psi_1},...,S_{m_1}^{\psi_1})$ and $(S_1^{\psi_2},...,S_{m_2}^{\psi_2})$. Notice that TSALBP design variables belong to a combinatorial optimization problem.
In order to calculate the similarity we first calculate the similarity of each station. A similarity index for each station $k$ in the two line configurations is given by Equation~\ref{eq:sim-station}.

\begin{equation} \label{eq:sim-station}
 Sim(S_k^{\psi_1},S_k^{\psi_2}) = \frac{2|S_k^{\psi_1} \cap S_k^{\psi_2}|}{|S_k^{\psi_1}| + |S_k^{\psi_2}|}.
\end{equation}

Taking into account the latter station similarity index, we define the similarity index $Sim \in [0,1]$ to measure how different two line configurations $\psi_1$ and $\psi_2$ are in $m$ stations, where $m=max\{m_1,m_2\}$ (Equation~\ref{eq:sim-T}). Two assembly line configurations are completely similar when $Sim(\psi_1, \psi_2) = 1$. 

\begin{equation} \label{eq:sim-T}
 Sim(\psi_1,\psi_2) = \frac{1}{m} \sum_{k=1}^m Sim(S_k^{\psi_1},S_k^{\psi_2}).
\end{equation}
 
\begin{figure*}[ht]
 \centering
\includegraphics[width=.55\textwidth]{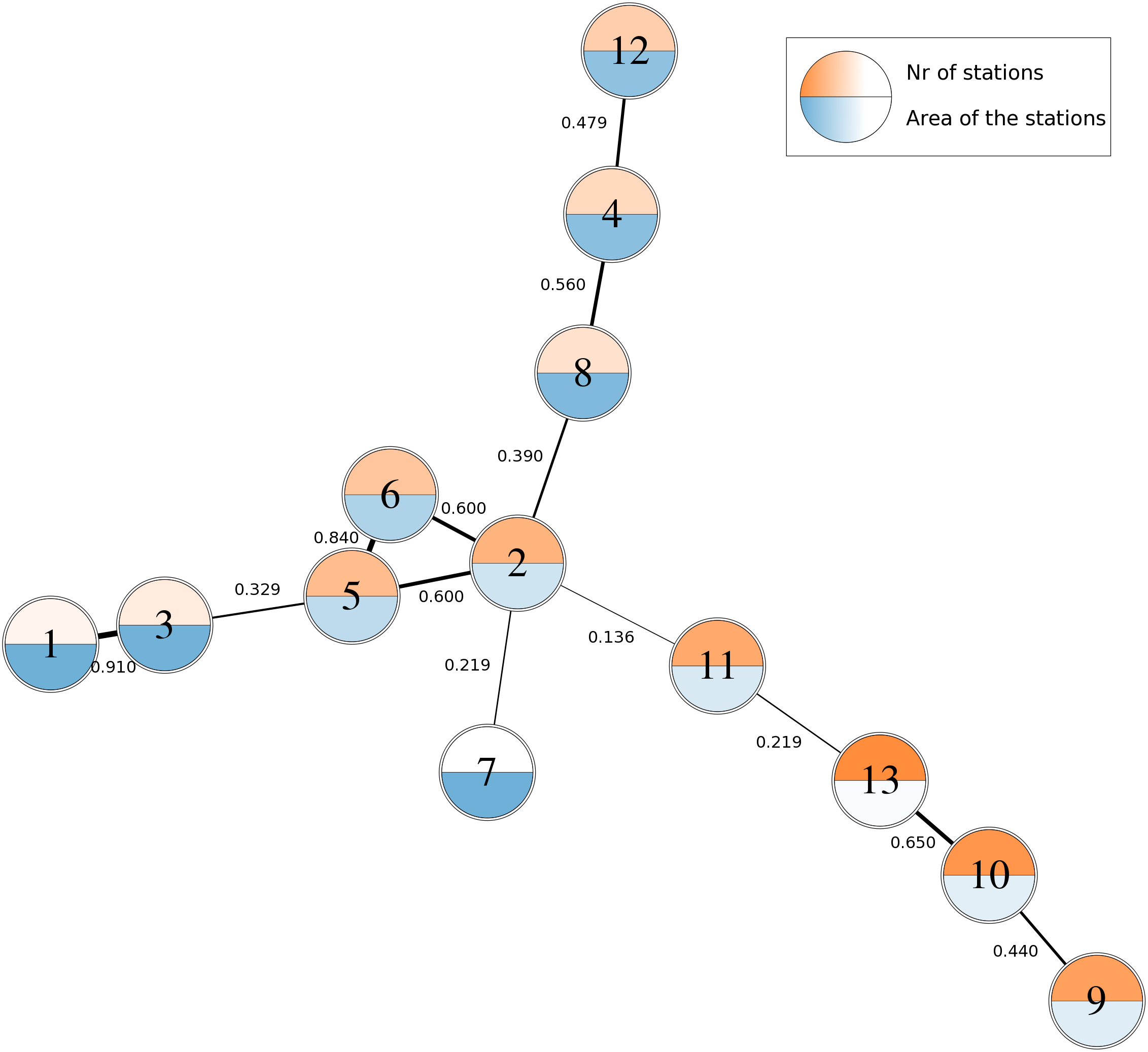}
\caption{\emph{moGram} generated for an instance of the TSALBP problem.}
\label{fig:tsalbp1}
\end{figure*}

\subsection{moGrams analysis and DM implications}

In Figure~\ref{fig:tsalbp1}, we can observe a \emph{moGram} generated for a TSALBP instance (420 tasks in a real automotive assembly line; see~\cite{Chica16OMEGA} for more details) with 13 solutions. The orange color represents the first objective, the number of stations, while the blue color shows the second objective, the area of the configuration line. Each edge has a label with its similarity value and its thickness depends on it.

The obtained \emph{moGram} allows to get the following observations from the visualization network:
\begin{itemize}
 \item The existing edges present very diverse similarity values, 0.91 being the highest value, whereas 0.14 being the lowest one. The highest values (0.91 and 0.84) are obtained between nodes 1, 3 and 5, 6, respectively. In contrast, the edge between nodes 2 and 11 (0.136) obtains the lowest value. The edges between nodes 2, 7 and 11, 13 also have low similarity value (0.219).
 \item We can distinguish three sub-networks based on the high similarity values between some pairs of nodes in the network. That happens at the top of the network (nodes 4, 8 and 12), at the bottom-right corner (nodes 9, 10 and 13) and also in the centre (nodes 2, 5, and 6). 
 \item A sub-network considering nodes 2, 5 and 6 (in the centre) is an interesting case. It is a fully connected sub-network with strong connections. The nodes 5 and 6 exhibit the highest similarity value (0.84), while their edges with node 2 have the same similarity (0.60).
\end{itemize}

From the \emph{moGram}, a decisor can get important insights about the TSALBP solving. For instance, the following items are problem-specific conclussions drawn from the visualization:

\begin{itemize}
 \item Node 13 exhibits the lowest number of stations (represented by a fully opaque orange color). According to its edges, this solution can be transformed into the solution 10 and also 11. Notice that transition to node 11 is more costly, due to the lower similarity value. 
 
 \item Solutions 1 and 7 characterize the lowest area (second objective reaches a fully transparent value). The former is well-connected to its neighbor (the thick edge towards node 3) demonstrating very similar characteristics in the design space. On the other hand, solution 7 is directly connected with solution 2 with contrasting objective values. 
 
 \item The most central node is the solution 2. It can be easily transformed to solutions 5 and 6; and it is also linked to other nodes (nodes 7, 8 and 11) with different characteristics in the objective space. Although the similarity values between solution 2 and the latter nodes are low, it is very useful information. It shows the flexibility of the solution 2 if the configuration line needed to be transformed to put more emphasis in the other objective.
\end{itemize}

\section{Application case 2: classifier ensembles}
\label{sec:appcase2}

\subsection{Problem description}

\begin{figure*}[ht]
 \centering
\includegraphics[width=.55\textwidth]{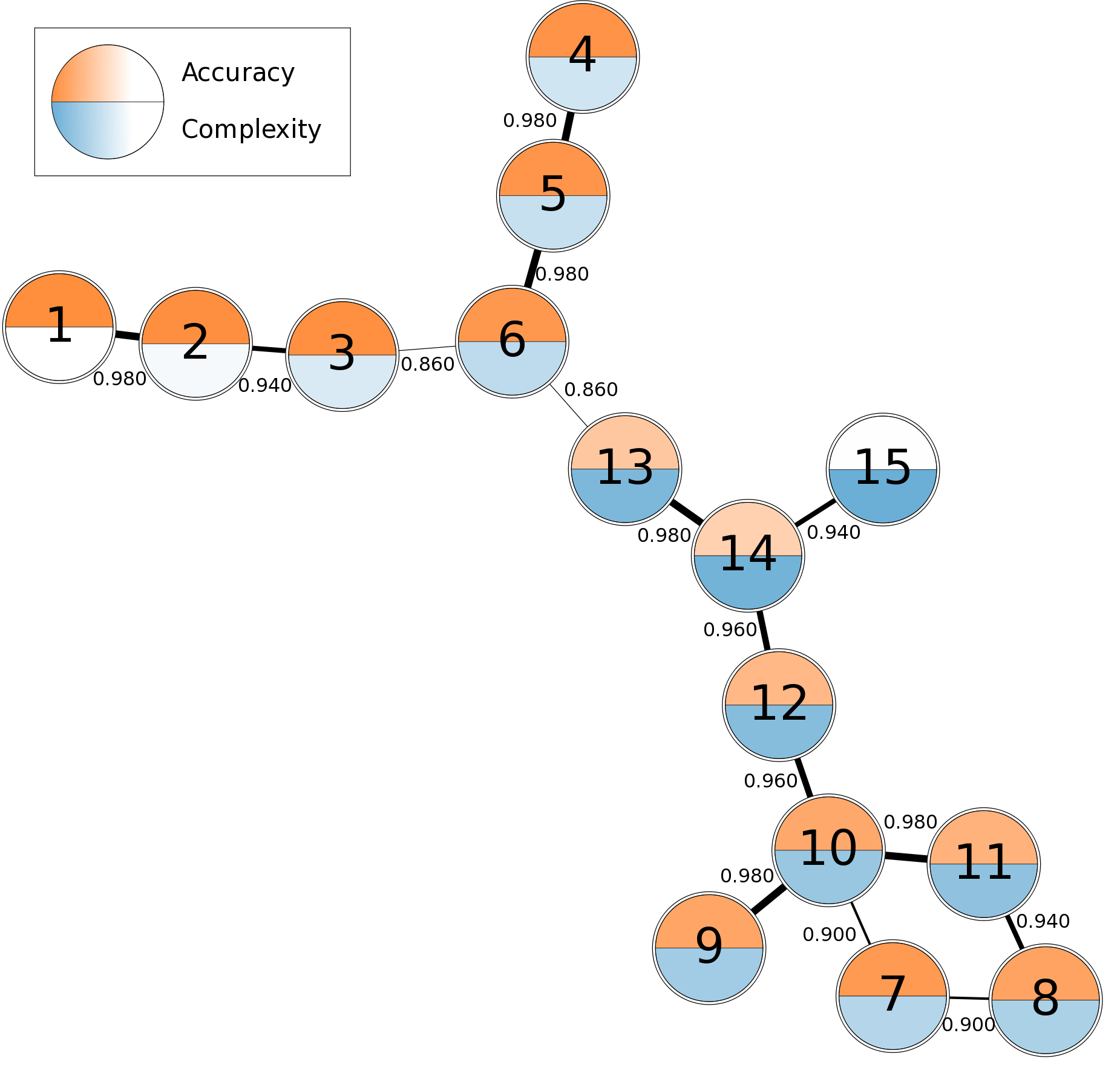}
\caption{moGram generated for the CE problem.}
\label{fig:ensembles1}
\end{figure*}

Classifier ensembles (CEs), also called multiclassification systems, are machine learning tools capable to obtain better performance than a single
classifier when dealing with complex classification problems, especially when the number of dimensions or the size of the data are really large \cite{Kun04}.

The overproduce-and-choose strategy (OCS)~\cite{Partridge96} (also known as test-and-select methodology \cite{Sharkey00}) is based on the generation of a large number of component classifiers, and a subsequent selection (removing duplicates and poor-performing candidate classifiers) to extract the best performing subset which composes the final CE. 

OCS methods aim at determining the optimal ensemble size by considering a trade-off between accuracy and complexity. 
The MO nature of this problem led researchers to use EMO algorithms and combine different measures in order to improve the accuracy-complexity trade-off of the final CE~\cite{Trawinski11,Trawinski2013,Oliveira05,Santos08}. In this application case, we focus on the EMO OCS approach based on two basic conflicting objectives, accuracy and complexity~\cite{Trawinski11}. Accuracy is defined as the proportion of correctly classified examples among the total number of examples, while complexity is represented by the number of classifiers.

From the DM point of view, it is of huge interest to obtain flexible CEs. Depending on the current demands, they can be converted to either a more accurate or to a less complex system by just adding or removing base classifiers. 
In an environment with limited resources (e.g. limited memory), the DM is interested in a model with the lowest complexity. When the correct answer is of big importance (e.g. breast cancer classification), the DM will choose the most accurate model. Beyond the information provided by the Pareto front approximation, \emph{moGrams} can be applied to the non-dominated CE solutions obtained in order to visually analyze their relationship in the design space, i.e. flexibility, centrality, among others.

The similarity metric uses information about the base classifiers composing the CE and is computed as follows. Two CE solutions $ce_1$ and $ce_2$ contain a subset of base classifiers from the initial pool of base classifiers (their total number is $n_{cl}$). These solutions are coded in binary strings $str^{ce_1}$ and $str^{ce_2}$ in a way that a binary digit/value $cl_k$ is assigned to each classifier that $str^{ce_1} = (cl_{1}^{ce_1}, cl_{2}^{ce_1}, ..., cl_{n_{cl}}^{ce_1})$ and $str^{ce_2} = (cl_{1}^{ce_2}, cl_{2}^{ce_2}, ..., cl_{n_{cl}}^{ce_2})$, respectively. When the value of $cl_k$ is 1, it means that a given classifier is included in the final ensemble, while 0 stands for classifier exclusion. We use the normalized \emph{Hamming distance}~\cite{hamming50} to compute the distance between two binary strings $str^{ce_1}$ and $str^{ce_2}$ (see Equation~\ref{eq:HamDist}).  

\begin{equation}
\label{eq:HamDist}
Hdist_{ce}(ce_1,ce_2) = \frac{\sum_{k=1}^{n_{cl}} |cl_{k}^{ce_1} - cl_{k}^{ce_2}| }{n_{cl}}
\end{equation}

Then, the distance metric $Hdist_{ce}(ce_1,ce_2)$ is subtracted from 1 in order to have a similarity measure $Sim(ce_1,ce_2) \in [0,1]$ (see Equation~\ref{eq:SimHamDist}). 

\begin{equation}
\label{eq:SimHamDist}
 Sim(ce_1,ce_2) = 1 - Hdist_{ce}(ce_1,ce_2)
\end{equation}

\subsection{moGrams analysis and DM implications}

Following the same procedure described in Section~\ref{sec:mograms} a \emph{moGram} is generated for an instance of a CE problem (the abalone dataset; see~\cite{Trawinski11} for more details) with 15 different solutions from the Pareto front approximation (Figure~\ref{fig:ensembles1}). The orange color is associated with the accuracy, while the blue color represents complexity. Since the edges present akin similarity values (their range is quite narrow), the final \emph{moGram} visualization was previously adjusted by the user in order to represent similarities in a clear way. 

The generated \emph{moGram} offers a clear representation of the solutions in the design space and leads to the following observations. Unlike the previous application case (see Section~\ref{sec:appcase1}), the existing edges present high similarity values (all above 0.86).  
When looking for the most accurate solution, node 1 is the suggested choice. However, it is separated from the rest of the nodes in the design space. So, if a flexible but still accurate solution is desired, then solution 6 is the proposed one, as it is connected to solutions with different objective values and can be easily modified (high similarity values) to more accurate or less complex solution (Solutions 3, 5 and 15).
In contrast, node 15 is the one having the less complex CE, but it is also isolated from the other nodes, as it has only one edge. Then, it is not the best choice when looking for flexibility.

Globally, the most flexible solutions are solutions 13 and 14, located in the centre of the network. They can be easily transformed to the other solution with different characteristics in the objective space. 
Finally, when looking for an accuracy-complexity trade-off solution, solutions 7 to 14 demonstrate these characteristics. Among them, solution 10 can be distinguished, as the most flexible one because it has the highest number of edges.

\section{Application case 3: Boolean queries for information retrieval}
\label{sec:appcase3}

\subsection{Problem description}

\begin{figure*}[htb!]
 \centering
\includegraphics[width=.65\textwidth]{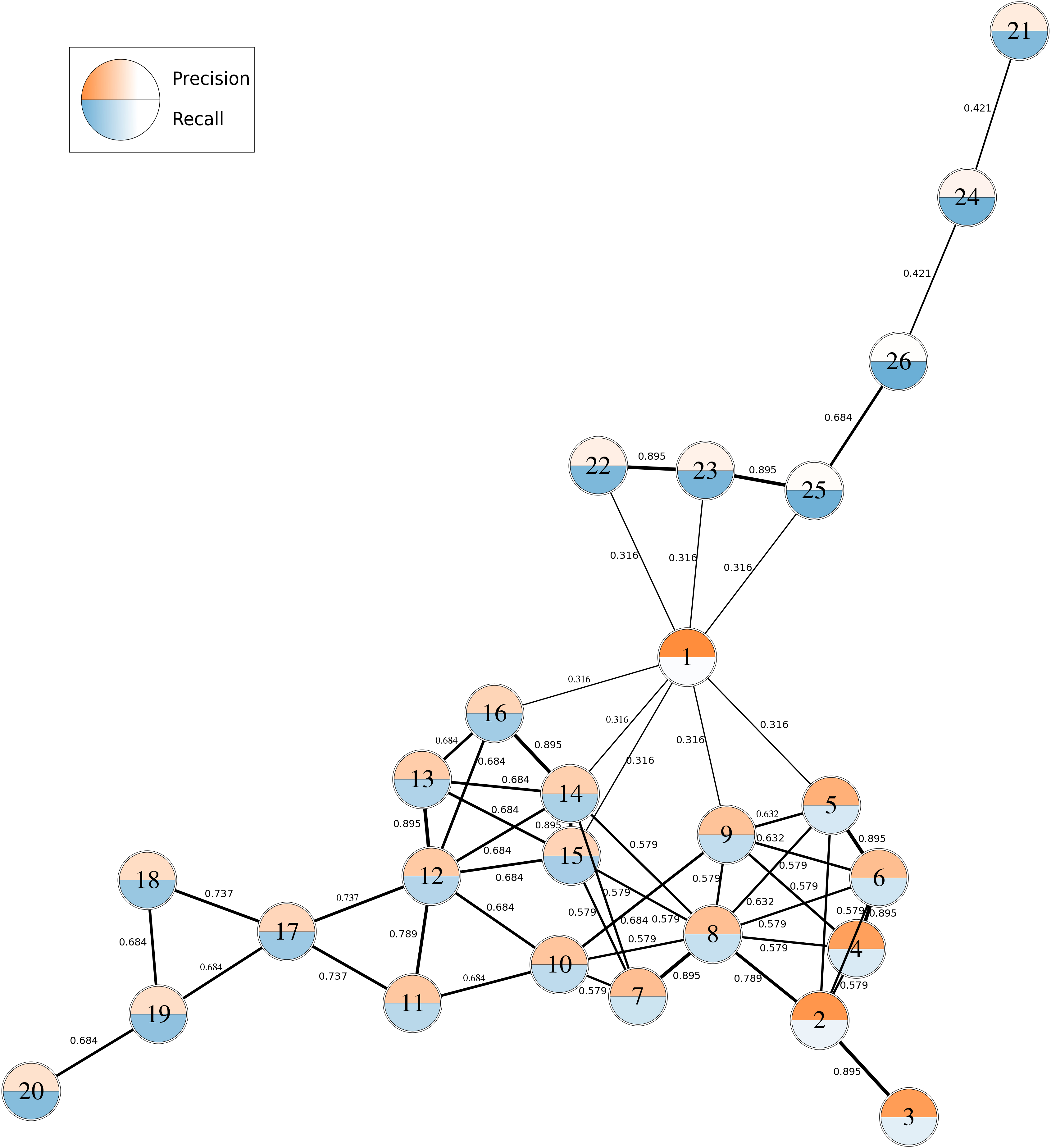}
\caption{\emph{moGram} generated for the Boolean queries problem.}
\label{fig:booleanQueries}
\end{figure*}

Information retrieval (IR) can be defined as the problem of selecting documentary information from a repository based on search queries introduced by a  user~\cite{Salton86,Baeza-Yates99}.
%
In Inductive Query By Example (IQBE)~\cite{Chen98}, a query is automatically generated, which describes the information contents of a set of documents provided by a user. Thus, it can be useful to assist the user in the query formulation process. The two most common criteria to measure the quality of IR system are precision and recall~\cite{Rijsbergen79}.
Boolean queries are defined by a query language, which is composed of query terms and it is combined with logical operators AND, OR and NOT~\cite{Rijsbergen79}.

As precision and recall are two conflicting criteria, the IQBE problem has a MO nature. In~\cite{Cordon2006,Herrera09a,Herrera09b}, the authors formulated IQBE as a MO problem and tackled it using EMO algorithms, obtaining several queries with a different precision-recall trade-off.

Precision is a measure of exactness or quality, while recall measures completeness or quantity. Thus, the DM might be interested in obtaining either more relevant documents (high precision) or most of the relevant documents from the repository (high recall). The information about the flexibility of the solution is valuable for the DM before choosing the final solution. Therefore, \emph{moGrams} can visually represent the mentioned flexibility of the solutions (i.e. how many changes are necessary to do in order to transform from a given solution to another).

Having two boolean queries $bq_1$ and $bq_2$ we consider them as strings in such a way that each term or operator of the boolean query is treated as a single entity/character $en_{i1}$ and $en_{j2}$ of the string, respectively. The similarity metric used is the \emph{edit} or \emph{Levenshtein distance}~\cite{levenshtein66} which measures the similarity between the generated boolean queries. \emph{Levenshtein distance} aims at finding the cheapest way to transform one string into another. Transformations in \emph{Levenshtein distance} are one-step based operations of insertion, deletion and substitution.  Then, \emph{Levenshtein distance} is computed by a dynamic programming algorithm and it is defined by the recurrence of Equation~\ref{eq:LevDist} (see~\cite{Navarro01} for more details).


\begin{equation}
Ldist(i,j) = \min \begin{cases}
		Ldist(i-1,j)+1 \\
		Ldist(i,j-1)+1 \\
		Ldist(i-1,j-1)+1_{(en_{i1} \ne en_{j2})} \\
             \end{cases}
\end{equation}
%

\noindent where $1_{(en_{i1} \ne en_{j2})}$ is the indicator function equal to 0 when $en_{i1}= en_{j2}$ and equal to 1 otherwise, while $i$ and $j$ are lengths of two compared boolean queries.

Then, the similarity metric $Sim(bq_1,bq_2)$ is normalized and subtracted from 1 in order to reflect similarity instead of the distance as presented in Equation~\ref{eq:SimHamDist}.

\begin{equation}
\label{eq:LevDist}
  Sim(bq_1,bq_2) = 1 - \frac{Ldist(|bq_1|, |bq_2|)}{max(|bq_1|,|bq_2|)} 
\end{equation}

\noindent with $|bq_1|$, $|bq_2|$  being the number of entities in the boolean queries $bq_1$ and $bq_2$, respectively.

\subsection{moGrams analysis and DM implications}

Fig.~\ref{fig:booleanQueries} shows the \emph{moGram} constructed for the IQBE problem (the Cranfield collection; see~\cite{Cordon2006} for more details). The orange and blue colors of nodes reflect precision and recall, respectively. In the light of this \emph{moGram}, we can observe that:

\begin{itemize}
\item We can uncover three subset of solutions at design space by observing the graphical representation:
  \begin{itemize}
  \item Solution 1, which maximizes precision, is far from the rest of solutions conforming the first subset. Although it has 7 edges those values are very low in all the cases.
  \item Solutions 2 to 20 are highly related each other and form a subset of solutions with good trade-off between precision and recall but without extreme values (except for solution 2). 
  \item Solutions 21 to 26 clearly maximize precision at the expense of reducing recall.
  \end{itemize}
\item Although solution 1 obtains the best precision, alternatively solution 2 also reaches very good precision values and is much more flexible, highly connected with solutions 4, 5, 6 and 8. In case the flexibility is required, solution 2 is suggested when looking for high precision.
\item If we look for a high recall, we can choose any of the solutions from 21 to 26. However, solution 25 obtains high recall and it is flexible, therefore it is one of the best options.
\item Globally, the most flexible solution is 8 as it has connections to nine other alternatives. However, any of the nodes from 2 to 20 are also highly flexible, to be changed to other solutions.
\end{itemize}


\section{Concluding remarks and future works}
\label{conclusions}

In this paper, we have proposed a novel methodology for visualizing and analyzing non-dominated solutions of MO problems. It provides the DM with a graphical exploration of the design space. In particular, it shows the relationships between the obtained solutions using a network. The so-called \emph{moGram} presents several interesting characteristics: it gives a clear insight into the design space, it incorporates jointly visualization of both design and objective spaces, and it is a generic and scalable approach. While generating the visualization, the DM interaction is also possible. Our proposal is based on SNA techniques for constructing, scaling and drawing the network. 

The \emph{moGrams} methodology was applied to three MO problems coming from very different research fields such as time and space assembly line balancing, classifier ensembles and boolean queries for information retrieval. We have shown the capabilities of \emph{moGrams} and how they can help the DM when choosing the final solution. Therefore, \emph{moGrams} is a powerful visualization technique, which aids in the understanding of the problem and the similarities between the solutions. Additionally, groups of solutions or the most flexible ones are easily detected thanks to this network-based visualization.

There are several future research lines that can be investigated. First, we can apply SNA techniques to analyze \emph{moGrams} in order to find the most significant nodes by using centrality measures such as centrality degree, closeness and betweenness for the nodes. Another opportunity is to apply community discovery algorithms to obtain clusters of solutions. We will also investigate the use of classical data mining techniques to obtain associations between the design space and the objective space for all the solutions of a problem. Finally, we would like to create an open Web tool for the generation and analysis of \emph{moGrams}.



\bibliographystyle{IEEEtran}
\bibliography{robustness,preferences,my_papers,mo_visualization,scientograms,ensembles,boolean_queries_ir,tsalbp}

\begin{thebibliography}{10}
\providecommand{\url}[1]{#1}
\csname url@rmstyle\endcsname
\providecommand{\newblock}{\relax}
\providecommand{\bibinfo}[2]{#2}
\providecommand\BIBentrySTDinterwordspacing{\spaceskip=0pt\relax}
\providecommand\BIBentryALTinterwordstretchfactor{4}
\providecommand\BIBentryALTinterwordspacing{\spaceskip=\fontdimen2\font plus
\BIBentryALTinterwordstretchfactor\fontdimen3\font minus
  \fontdimen4\font\relax}
\providecommand\BIBforeignlanguage[2]{{%
\expandafter\ifx\csname l@#1\endcsname\relax
\typeout{** WARNING: IEEEtran.bst: No hyphenation pattern has been}%
\typeout{** loaded for the language `#1'. Using the pattern for}%
\typeout{** the default language instead.}%
\else
\language=\csname l@#1\endcsname
\fi
#2}}

\bibitem{Chankong83}
V.~Chankong and Y.~Y. Haimes, \emph{Multiobjective Decision Making Theory and
  Methodology}.\hskip 1em plus 0.5em minus 0.4em\relax North-Holland, 1983.

\bibitem{Coello07}
C.~A. Coello, G.~B. Lamont, and D.~A. {Van Veldhuizen}, \emph{Evolutionary
  Algorithms for Solving Multi-objective Problems (2nd edition)}.\hskip 1em
  plus 0.5em minus 0.4em\relax Springer, 2007.

\bibitem{Deb01}
K.~Deb, \emph{Multi-objective Optimization Using Evolutionary
  Algorithms}.\hskip 1em plus 0.5em minus 0.4em\relax Wiley, 2001.

\bibitem{Bonissone08}
P.~Bonissone, ``Research issues in multi criteria decision making ({MCDM}): The
  impact of uncertainty in solution evaluation,'' in \emph{12th International
  Conference on Processing and Management of Uncertainty in Knowledge-based
  Systems ({IPMU})}, M\'{a}laga (Spain), June 2008, pp. 1409--1416.

\bibitem{Miettinen14}
K.~Miettinen, ``Survey of methods to visualize alternatives in multiple
  criteria decision making problems,'' \emph{OR Spectrum}, pp. 1--35, 2014.

\bibitem{Chica16OMEGA}
M.~Chica, J.~Bautista, {\'O}.~Cord{\'o}n, and S.~Damas, ``A multiobjective
  model and evolutionary algorithms for robust time and space assembly line
  balancing under uncertain demand,'' \emph{Omega}, vol.~58, pp. 55--68, 2016.

\bibitem{Fleming05}
P.~J. Fleming, R.~C. Purshouse, and R.~J. Lygoe, ``Many-objective optimization:
  An engineering design perspective,'' in \emph{Evolutionary Multi-Criterion
  Optimization}.\hskip 1em plus 0.5em minus 0.4em\relax Springer, 2005, pp.
  14--32.

\bibitem{Stump03}
G.~M. Stump, M.~Yukish, T.~W. Simpson, and E.~N. Harris, ``Design space
  visualization and its application to a design by shopping paradigm,'' in
  \emph{ASME Design Engineering Technical Conferences-Design Automation
  Conference}.\hskip 1em plus 0.5em minus 0.4em\relax Chicago, IL, ASME, Paper
  No. DETC2003/DAC-48785, 2003.

\bibitem{Walker13}
D.~Walker, R.~Everson, and J.~Fieldsend, ``Visualizing mutually nondominating
  solution sets in many-objective optimization,'' \emph{Evolutionary
  Computation, IEEE Transactions on}, vol.~17, no.~2, pp. 165--184, April 2013.

\bibitem{Tusar15}
T.~Tusar and B.~Filipic, ``Visualization of pareto front approximations in
  evolutionary multiobjective optimization: A critical review and the
  prosection method,'' \emph{Evolutionary Computation, IEEE Transactions on},
  vol.~19, no.~2, pp. 225--245, April 2015.

\bibitem{Lotov04}
A.~V. Lotov, V.~A. Bushenkov, and G.~K. Kamenev, \emph{Interactive decision
  maps: Approximation and visualization of Pareto frontier}.\hskip 1em plus
  0.5em minus 0.4em\relax Springer, 2004, vol.~89.

\bibitem{Kollat07}
J.~B. Kollat and P.~Reed, ``A framework for visually interactive
  decision-making and design using evolutionary multi-objective optimization
  ({VIDEO}),'' \emph{Environmental Modelling \& Software}, vol.~22, no.~12, pp.
  1691--1704, 2007.

\bibitem{Kurasova13}
O.~Kurasova, T.~Petkus, and E.~Filatovas, ``Visualization of pareto front
  points when solving multi-objective optimization problems,''
  \emph{Information Technology And Control}, vol.~42, no.~4, pp. 353--361,
  2013.

\bibitem{Obayashi05}
S.~Obayashi, S.~Jeong, and K.~Chiba, ``Multi-objective design exploration for
  aerodynamic configurations,'' in \emph{Proceedings of the AIAA Fluid dynamics
  conference and exhibit}, vol. 4666, 2005.

\bibitem{Pryke07}
A.~Pryke, S.~Mostaghim, and A.~Nazemi, ``Heatmap visualization of population
  based multi objective algorithms,'' in \emph{Evolutionary Multi-Criterion
  Optimization}, ser. Lecture Notes in Computer Science, S.~Obayashi, K.~Deb,
  C.~Poloni, T.~Hiroyasu, and T.~Murata, Eds.\hskip 1em plus 0.5em minus
  0.4em\relax Springer Berlin Heidelberg, 2007, vol. 4403, pp. 361--375.

\bibitem{Eddy02}
J.~Eddy and K.~Lewis, ``Visualization of multidimensional design and
  optimization data using cloud visualization,'' in \emph{ASME Design Technical
  Conferences, Design Automation Conference}, 2002.

\bibitem{Agrawal04}
G.~Agrawal, K.~Lewis, K.~Chugh, C.~Huang, S.~Parashar, and C.~Bloebaum,
  ``Intuitive visualization of pareto frontier for multi-objective optimization
  in n-dimensional performance space,'' in \emph{10th AIAA/ISSMO
  multidisciplinary analysis and optimization conference}, 2004, pp.
  1523--1533.

\bibitem{Scott2000}
J.~Scott, \emph{Social Network Analysis: A Handbook (2nd edition)}.\hskip 1em
  plus 0.5em minus 0.4em\relax Sage Publications, 2000.

\bibitem{wasserman1994social}
S.~Wasserman and K.~Faust, \emph{Social Network Analysis: Methods And
  Applications (Structural Analysis in the Social Sciences)}.\hskip 1em plus
  0.5em minus 0.4em\relax Cambridge University Press, 1994.

\bibitem{Dearholt90}
D.~W. Dearholt and R.~W. Schvaneveldt, ``Properties of pathfinder networks,''
  in \emph{Pathfinder associative networks: Studies in knowledge organization},
  R.~Schvaneveldt, Ed.\hskip 1em plus 0.5em minus 0.4em\relax Ablex Publishing
  Corporation, 1990, pp. 1--30.

\bibitem{Sch89}
R.~W. Schvaneveldt, F.~T. Durso, and D.~W. Dearholt, ``Network structures in
  proximity data,'' \emph{The psychology of learning and motivation: Advances
  in research and theory}, vol.~24, pp. 249--284, 1989.

\bibitem{MoyaAnegon07}
F.~Moya-Aneg{\'o}n, B.~Vargas-Quesada, Z.~Chinchilla-Rodr{\'i}guez,
  E.~Corera-{\'A}lvarez, F.~J. Mu{\~n}oz-Fern{\'a}ndez, and V.~Herrero-Solana,
  ``Visualizing the marrow of science,'' \emph{Journal of the American Society
  for Information Science and Technology}, vol.~58, no.~14, pp. 2167--2179,
  2007.

\bibitem{Chica10Ins}
M.~Chica, O.~Cord{\'o}n, S.~Damas, and J.~Bautista, ``Multiobjective,
  constructive heuristics for the 1/3 variant of the time and space assembly
  line balancing problem: {ACO} and random greedy search,'' \emph{Information
  Sciences}, vol. 180, pp. 3465--3487, 2010.

\bibitem{Chica11CAIE}
M.~Chica, O.~Cord{\'o}n, and S.~Damas, ``An advanced multi-objective genetic
  algorithm design for the time and space assembly line balancing problem,''
  \emph{Computers and Industrial Engineering}, vol.~61, no.~1, pp. 103--117,
  2011.

\bibitem{Partridge96}
D.~Partridge and W.~Yates, ``Engineering multiversion neural-net systems,''
  \emph{Neural Computation}, vol.~8, no.~4, pp. 869--893, 1996.

\bibitem{Kun04}
L.~Kuncheva, \emph{Combining Pattern Classifiers: Methods and
  Algorithms}.\hskip 1em plus 0.5em minus 0.4em\relax Wiley, 2004.

\bibitem{Trawinski11}
K.~Trawi\'nski, A.~Quirin, and O.~Cord{\'o}n, ``A study on the use of
  multi-objective genetic algorithms for classifier selection in furia-based
  fuzzy multiclassifers,'' \emph{International Journal of Computational
  Intelligence Systems}, vol.~5, no.~2, pp. 231--253, 2012.

\bibitem{Trawinski2013}
K.~Trawi\'nski, O.~Cordón, A.~Quirin, and L.~Sánchez, ``Multiobjective genetic
  classifier selection for random oracles fuzzy rule-based classifier
  ensembles: How beneficial is the additional diversity?''
  \emph{Knowledge-Based Systems}, vol.~54, pp. 3 -- 21, 2013.

\bibitem{Cordon2006}
O.~Cord\'on, E.~Herrera-Viedma, and M.~Luque, ``Improving the learning of
  boolean queries by means of a multiobjective \{IQBE\} evolutionary
  algorithm,'' \emph{Information Processing and Management}, vol.~42, no.~3,
  pp. 615 -- 632, 2006.

\bibitem{Chambers82}
J.~M. Chambers and B.~Kleiner, ``10 graphical techniques for multivariate data
  and for clustering,'' in \emph{Classification Pattern Recognition and
  Reduction of Dimensionality}, ser. Handbook of Statistics, P.~Krishnaiah and
  L.~Kanal, Eds.\hskip 1em plus 0.5em minus 0.4em\relax Elsevier, 1982, vol.~2,
  pp. 209 -- 244.

\bibitem{inselberg09}
A.~Inselberg, \emph{Parallel Coordinates: Visual Multidimensional Geometry and
  Its Applications}, ser. Advanced series in agricultural sciences.\hskip 1em
  plus 0.5em minus 0.4em\relax Springer New York, 2009.

\bibitem{Winer02}
E.~Winer and C.~Bloebaum, ``Development of visual design steering as an aid in
  large-scale multidisciplinary design optimization. part ii: method
  validation,'' \emph{Structural and Multidisciplinary Optimization}, vol.~23,
  no.~6, pp. 425--435, 2002.

\bibitem{Agrawal06}
G.~Agrawal, K.~Lewis, and C.~Bloebaum, ``Intuitive visualization of hyperspace
  pareto frontier,'' in \emph{44th AIAA Aerospace Sciences Meeting and
  Exhibit}, 2006, pp. 8783--8796.

\bibitem{Jeong05}
S.~Jeong, K.~Chiba, and S.~Obayashi, ``Data mining for aerodynamic design
  space,'' \emph{Journal of aerospace computing, information, and
  communication}, vol.~2, no.~11, pp. 452--469, 2005.

\bibitem{Jeong07}
M.-J. Jeong, T.~Kobayashi, and S.~Yoshimura, ``Multidimensional visualization
  and clustering for multiobjective optimization of artificial satellite heat
  pipe design,'' \emph{Journal of mechanical science and technology}, vol.~21,
  no.~12, pp. 1964--1972, 2007.

\bibitem{Blasco08}
X.~Blasco, J.~M. Herrero, J.~Sanch\'{i}s, and M.~Mart\'{i}nez, ``A new
  graphical visualization of n-dimensional {Pareto} front for decision-making
  in multiobjective optimization,'' \emph{Information Sciences}, vol. 178,
  no.~20, pp. 3908 -- 3924, 2008.

\bibitem{Kubota14}
M.~Kubota, T.~Itoh, S.~Obayashi, and Y.~Takeshima, ``{EVOLVE}: A visualization
  tool for multi-objective optimization featuring linked view of explanatory
  variables and objective functions,'' in \emph{Information Visualisation (IV),
  2014 18th International Conference on}, July 2014, pp. 351--356.

\bibitem{PanchoIEEETFS2013}
D.~P. Pancho, J.~M. Alonso, O.~Cord\'{o}n, A.~Quirin, and L.~Magdalena,
  ``{FINGRAMS}: visual representations of fuzzy rule-based inference for expert
  analysis of comprehensibility,'' \emph{IEEE Transactions on Fuzzy Systems},
  vol.~21, no.~6, pp. 1133--1149, 2013.

\bibitem{Vargas2007visualizing}
B.~Vargas-Quesada and F.~Moya-Aneg\'{o}n, \emph{Visualizing the structure of
  science}.\hskip 1em plus 0.5em minus 0.4em\relax Springer-Verlag, 2007.

\bibitem{chen2003visualizing}
C.~Chen and S.~Morris, ``Visualizing evolving networks: Minimum spanning trees
  versus pathfinder networks,'' in \emph{IEEE Symposium on Information
  Visualization}, 2003, pp. 67--74.

\bibitem{zizi1994accessing}
M.~Zizi and M.~Beaudouin-Lafon, ``Accessing hyperdocuments through interactive
  dynamic maps,'' in \emph{Proceedings of the ACM European conference on
  Hypermedia technology}, 1994, pp. 126--135.

\bibitem{noel2002visualization}
S.~Noel, C.~H. Chu, and V.~Raghavan, ``Visualization of document co-citation
  counts,'' in \emph{IEEE Symposium on Information Visualisation}, 2002, pp.
  691--696.

\bibitem{battista1999graph}
G.~di~Battista, P.~Eades, R.~Tamassia, and I.~Tollis, \emph{Graph drawing:
  algorithms for the visualization of graphs}.\hskip 1em plus 0.5em minus
  0.4em\relax Upper Saddle River, N.J: Prentice Hall, 1998.

\bibitem{kobourov2005force}
S.~G. Kobourov, ``Force-directed drawing algorithms,'' in \emph{Handbook of
  Graph Drawing and Visualization}, R.~Tamassia, Ed.\hskip 1em plus 0.5em minus
  0.4em\relax CRC Press, 2012, ch.~12.

\bibitem{KamadaKawai1989}
T.~Kamada and S.~Kawai, ``An algorithm for drawing general undirected graphs,''
  \emph{Information Processing Letters}, vol.~31, no.~1, pp. 7--15, 1989.

\bibitem{fruchterman1991graph}
T.~M.~J. Fruchterman and E.~M. Reingold, ``Graph drawing by force-directed
  placement,'' \emph{Software: Practice and experience}, vol.~21, no.~11, pp.
  1129--1164, 1991.

\bibitem{Graphviz1999}
E.~R. Gansner and S.~C.North, ``An open graph visualization system and its
  applications to software engineering,'' \emph{Software: Practice and
  Experience}, vol.~30, no.~11, pp. 1203--1233, 1999.

\bibitem{Battaia13}
O.~Batta{\"i}a and A.~Dolgui, ``A taxonomy of line balancing problems and their
  solution approaches,'' \emph{International Journal of Production Economics},
  vol. 142, no.~2, pp. 259--277, 2013.

\bibitem{Chica12EngAppAI}
M.~Chica, O.~Cord{\'o}n, S.~Damas, and J.~Bautista, ``Multiobjective memetic
  algorithms for time and space assembly line balancing,'' \emph{Engineering
  Applications of Artificial Intelligence}, vol.~25, no.~2, pp. 254--273, 2012.

\bibitem{Chica13IJPE}
------, ``A robustness information and visualization model for time and space
  assembly line balancing under uncertain demand,'' \emph{International Journal
  of Production Economics}, vol. 145, pp. 761--772, 2013.

\bibitem{Sharkey00}
A.~Sharkey and N.~Sharkey, ``The {\it test and select} approach to ensemble
  combination,'' in \emph{International Workshop on Multiclassifier Systems},
  Cagliari (Italy), 2000, pp. 30--44.

\bibitem{Oliveira05}
L.~Oliveira, M.~Morita, R.~Sabourin, and F.~Bortolozzi, ``Multi-objective
  genetic algorithms to create ensemble of classifiers,'' \emph{Lecture Notes
  in Computer Science}, vol. 3410, pp. 592--606, 2005.

\bibitem{Santos08}
E.~D. Santos, R.~Sabourin, and P.~Maupin, ``A dynamic overproduce-and-choose
  strategy for the selection of classifier ensembles,'' \emph{Pattern
  Recognition}, vol.~41, no.~10, pp. 2993--3009, 2008.

\bibitem{hamming50}
R.~Hamming, ``{Error Detecting and Error Correcting Codes},'' \emph{Bell System
  Techincal Journal}, vol.~29, pp. 147--160, 1950.

\bibitem{Salton86}
G.~Salton and M.~J. McGill, \emph{Introduction to Modern Information
  Retrieval}.\hskip 1em plus 0.5em minus 0.4em\relax McGraw-Hill, Inc., 1986.

\bibitem{Baeza-Yates99}
R.~A. Baeza-Yates and B.~Ribeiro-Neto, \emph{Modern Information
  Retrieval}.\hskip 1em plus 0.5em minus 0.4em\relax Addison-Wesley Longman
  Publishing Co., Inc., 1999.

\bibitem{Chen98}
H.~Chen, G.~Shankaranarayanan, L.~She, and A.~Iyer, ``A machine learning
  approach to inductive query by,'' \emph{Journal of the American Society for
  Information Science}, vol.~49, pp. 693--705, 1998.

\bibitem{Rijsbergen79}
C.~J.~V. Rijsbergen, \emph{Information Retrieval}, 2nd~ed.\hskip 1em plus 0.5em
  minus 0.4em\relax Butterworth-Heinemann, 1979.

\bibitem{Herrera09a}
A.~G. L\'{o}pez-Herrera, E.~Herrera-Viedma, and F.~Herrera, ``A study of the
  use of multi-objective evolutionary algorithms to learn boolean queries: A
  comparative study,'' \emph{Journal of the American Society for Information
  Science and Technology}, vol.~60, no.~6, pp. 1192--1207, 2009.

\bibitem{Herrera09b}
------, ``Applying multi-objective evolutionary algorithms to the automatic
  learning of extended boolean queries in fuzzy ordinal linguistic information
  retrieval systems,'' \emph{Fuzzy Sets and Systems}, vol. 160, no.~15, pp.
  2192--2205, 2009.

\bibitem{levenshtein66}
V.~I. Levenshtein, ``{Binary codes capable of correcting deletions, insertions,
  and reversals},'' \emph{Soviet Physics Doklady}, vol.~10, no.~8, pp.
  707--710, 1966.

\bibitem{Navarro01}
G.~Navarro, ``A guided tour to approximate string matching,'' \emph{ACM
  Computing Surveys}, vol.~33, no.~1, pp. 31--88, Mar. 2001.

\end{thebibliography}


\end{document}